\documentclass[aip,apl,twocolumn,11pt]{revtex4-1}
\pdfoutput=1
\usepackage{amsmath}
\usepackage{graphicx}
\usepackage{natbib}
\usepackage{mathrsfs}

\usepackage[T1]{fontenc}
\usepackage[letterpaper,textwidth=7in,top=.75in,bottom=.75in]{geometry}
\begin{document}
\title{Broadband magnetometry by infrared-absorption detection of diamond NV centers}

 \author{V.~M.~Acosta}
 \email{vmacosta@berkeley.edu}
    \address{
     Physics Department, University of California,
     Berkeley, CA 94720
    }
 \author{E.~Bauch}
    \address{
     Physics Department, University of California,
     Berkeley, CA 94720
    }
 \author{A.~Jarmola}
    \address{
     Physics Department, University of California,
     Berkeley, CA 94720
    }
 \author{L.~J.~Zipp}
    \address{
     Physics Department, University of California,
     Berkeley, CA 94720
    }
 \author{M.~P.~Ledbetter}
    \address{
     Physics Department, University of California,
     Berkeley, CA 94720
    }
 \author{D.~Budker}
    \address{
     Physics Department, University of California,
     Berkeley, CA 94720
    }
    \address{Nuclear Science Division, Lawrence Berkeley Laboratory,
     Berkeley CA 94720
    }
\date{\today}

%\begin{article}
\begin{abstract}
We demonstrate magnetometry by detection of the spin state of high-density nitrogen-vacancy (NV) ensembles in diamond using optical absorption at 1042 nm. With this technique, measurement contrast and collection efficiency can approach unity, leading to an increase in magnetic sensitivity compared to the more common method of collecting red fluorescence. Working at 75 K with a sensor with effective volume $50\times50\times300~{\rm \mu m}^3$, we project photon shot-noise limited sensitivity of $5~{\rm pT}$ in one second of acquisition and bandwidth from DC to a few MHz. Operation in a gradiometer configuration yields a noise floor of $7~{\rm nT_{rms}}$ at $\sim110~{\rm Hz}$ in one second of acquisition.
%07.55.Ge Magnetometers for magnetic field measurements, 61.72.jn Color centers
%76.30.Mi Color centers and other defects
%81.05.Uw  Carbon, diamond, graphite
\end{abstract}
%\pacs{(07.55.Ge) Magnetometers for magnetic field measurements; (61.72.jn) Color centers; (76.30.Mi) Color centers and other defects; (81.05.Uw) Carbon, diamond, graphite}
\maketitle
%At the core of quantum metrology is the ability to obtain narrow resonances, or long-lived coherences, in the ensemble of spins under study. Narrow resonances in alkali vapor cells \cite{BUD2007} have been used to measure magnetic fields with a better sensitivity \cite{KOM2003,DAN2009} than the previously most sensitive superconducting quantum interference devices (SQUIDs) \cite{CLA2004}. These so-called atomic magnetometers have been widely employed both in laboratory precision measurements \cite{MUR1989,BER1995,YOU1996,GAR2001,VAS2009,BRO2010} as well as field applications such as space physics \cite{DOU2005,HIG2009}, medical research \cite{BIS2003,XIA2006}, magnetic micro-particle detection \cite{XU2006}, and, recently, nuclear magnetic resonance (NMR) \cite{SAV2007,XU2006PNAS,DON2009} in microfluidic chips \cite{LED2008PNAS} even in the zero-field environment \cite{LED2009}.
%
%There are areas where the existing technologies and approaches find their limits. While atomic magnetometers can measure magnetic fields with exceptional sensitivity and without cryogenics, spin-altering collisions limit the sensitivity of small sensors \cite{SHA2007,MIK2009}. In order to probe pT-scale magnetic fields with micrometer resolution, measurements using SQUIDS \cite{CLA2004,JAM2001,CLE2006} and magneto-resistive devices \cite{TSY2001,PAN2004,VER2008} are commonly used. These solid-state sensors are widely available, but the former require cryogenic cooling and the latter suffer from 1/f noise \cite{STU2005}, which limits the range of possible applications.

Recently, a technique for measuring magnetic fields at the micro- and nanometer scale has emerged based on nitrogen-vacancy (NV) electron spin resonances in diamond \cite{BAL2008,MAZ2008NATURE}. This technique offers an exceptional combination of sensitivity and spatial resolution in a wide temperature range, from $0~{\rm K}$ to above room temperature. Sensors employing ensembles of NV centers promise the highest sensitivity \cite{TAY2008,ACO2009} and have recently been demonstrated \cite{BOU2010,STE2010,MAE2010}. These first-generation magnetometers have measured sub-micron-scale fields, but their sensitivity was limited by background fluorescence and poor photon collection efficiency.

In this Letter, we demonstrate a technique to read out the NV spin state using infrared (IR) absorption at 1042 nm \cite{ROG2008,ACO2010ARXIV}. With this technique, measurement contrast and collection efficiency can approach unity, leading to an overall increase in magnetic sensitivity. We perform measurements at 45-75 K on a sensor with active volume $\sim50\times50\times300~{\rm \mu m}^3$, revealing magnetic resonances with amplitude and width corresponding to a shot-noise-limited sensitivity of $5~{\rm pT}$ in one second of acquisition and a measurement bandwidth from DC up to a few MHz. We demonstrate operation in a gradiometer configuration, with a sensitivity of $7~{\rm nT_{rms}}$ in one second of acquisition at $\sim110~{\rm Hz}$ bandwidth, and outline a design for a room-temperature device employing a low-finesse optical cavity with sensitivity approaching the spin-projection noise limit.

Spin-based magnetometers are fundamentally limited by the quantum noise associated with spin projection. The minimum detectable magnetic field for a sample of spins with density $n$ in a volume $V$ is given by \cite{BUD2007,TAY2008}:
\begin{equation}
\label{eq:1}
\delta B_q \simeq \frac{1}{\gamma}\frac{1}{\sqrt{n V t_m T_2^{\ast}}},
\end{equation}
where $\gamma=1.761\times10^{11}~{\rm s^{-1} T^{-1}}$ is the NV gyromagnetic ratio \cite{LOU1978} and $T_2^{\ast}$ is the electron spin dephasing time, and $t_m\gtrsim T_2^{\ast}$ is the measurement time. The sample \cite{ACO2009} used in this work (labeled S2), contains an NV density of $n=7\times10^{17}~{\rm cm^{-3}}$ for each of the four crystallographic orientations and typically exhibits $T_2^{\ast}=0.15~{\rm \mu s}$. With optimal detection the magnetic sensitivity approaches the limit set by Eq. \eqref{eq:1}, and we find the noise floor in this sample of $\sim20~{\rm pT/\mu m^{3/2}}$ for $t_m=1~{\rm s}$, or $\sim10~{\rm fT}$ for the active volume used here.

However, reaching this level of sensitivity requires an improvement over the commonly used technique of detecting spin selective fluorescence \cite{VAN1988,JEL2006}. For sufficiently low measurement contrast, $R$ (relative difference in detected signal depending on spin-projection), the sensitivity using the fluorescence technique can be estimated \cite{TAY2008,ACO2009} by modifying Eq. \eqref{eq:1} as $\delta B_{fl}\approx\delta B_q/(R\sqrt{\eta})$, where $\eta$ is the detection efficiency. Recent experiments \cite{STE2010,MAE2010} yielded typical values of $R\sim0.03$ and $\eta\sim0.001$, making the best possible sensitivity, in the absence of excess technical noise or other broadening mechanisms, about three orders of magnitude worse than the spin-projection-noise limit. The contrast is limited by non-ideal branching ratios to the dark singlet states \cite{MAN2006,ROG2008} and high background fluorescence from defects which do not contribute to the magnetometer signal \cite{ALE2007}, while $\eta$ is limited by the field of view of the detection optics and sub-unity quantum efficiency of the detected transition \cite{HAN2010,BAB2010}.

\begin{figure}
\centering
  \includegraphics[width=0.45\textwidth]{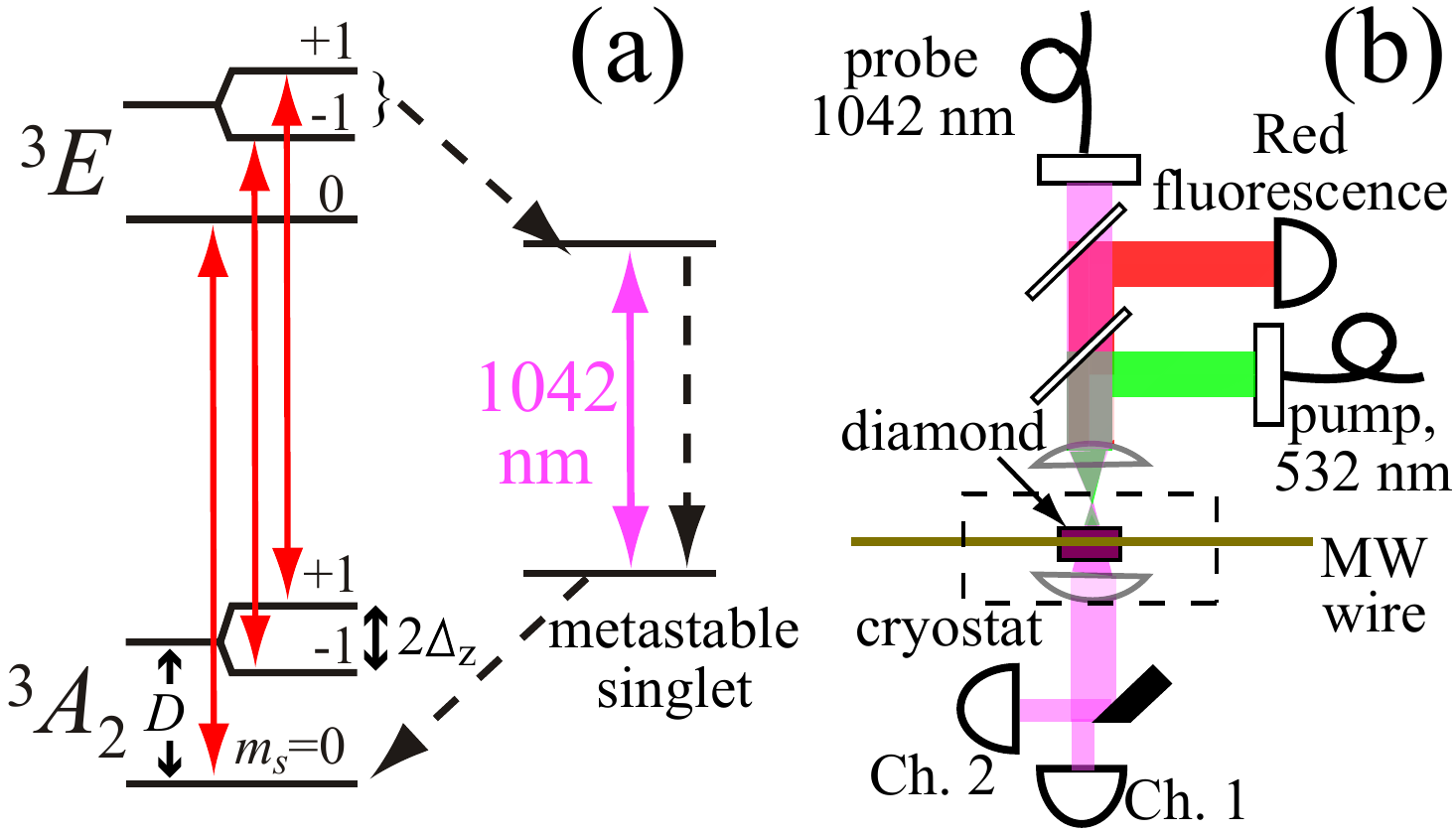}
\caption{\label{fig:magapp} \footnotesize (a) Level structure of the NV center and allowed optical transitions. Radiative (non-radiative) transitions are represented by solid (dashed) lines. (b) IR absorption gradiometer apparatus. The green pump and IR probe beams were focused to a diameter of $\sim30~{\rm \mu m}$ near the surface of the diamond, and two halves of the transmitted IR beam were detected with separate photodiodes. MW--microwave. }
\end{figure}

By using IR absorption detection, we circumvent most of these problems and achieve considerably higher sensitivity. Figure \ref{fig:magapp}(a) displays the level structure of the NV center with allowed radiative and non-radiative transitions. The center has a paramagnetic ($S=1$) ground state, with a zero-field splitting of $D\approx2.87~{\rm GHz}$ at room-temperature \cite{ACO2010}. At low magnetic field ($<<0.1~{\rm T}$), the magnetic sublevels shift by $\Delta_z\approx \gamma m_s B_z/(2\pi)$, where $B_z$ is the projection of the magnetic field along the NV axis. Optical pumping via a spin-selective decay path \cite{MAN2006,ROG2008,ACO2010ARXIV} involving a 1042 nm transition, populates NV centers in the $m_s=0$ ground-state sublevel. The same decay path is also responsible for the drop in fluorescence upon application of resonant microwaves, which is the principle of operation of recent magnetometry demonstrations \cite{BAL2008,MAZ2008NATURE,BAL2009,BOU2010,MAE2010,STE2010}.

The apparatus is illustrated in Fig. \ref{fig:magapp}(b). Pump (532 nm) and probe (1042 nm) beams were overlapped and focused to a waist of $\sim30~{\rm \mu m}$ diameter, approximately $0.5~{\rm mm}$ before the diamond surface. The diverging beam had a diameter of $\sim50~{\rm \mu m}$ as it passed through the diamond, and was sufficiently far from the focal point (more than one Rayleigh range) that it exhibited far-field diffraction. Fluorescence was collected via the same lens, spectrally filtered to pass 650-800 nm light, and detected by a photodiode. The transmitted probe beam was split in two halves by a sharp-edge mirror and detected with separate photodiodes. The diamond was housed in a liquid-helium cryostat (Janis ST-500) and microwaves were delivered via a wire placed $\sim2~{\rm mm}$ from the illuminated region. No magnetic shielding was used.

The principle behind our technique is the following: under continuous optical pumping, we detect the population of NV centers in the metastable singlet (MS) by monitoring the transmission of the 1042 nm probe beam and use this to read out the spin polarization of the ensemble. In the absence of resonant microwaves, NV centers are pumped into the $m_s=0$ ground-state sublevel and there is reduced population in the MS corresponding to maximum probe transmission. Under application of microwaves with frequency $D\pm\gamma B_{z_i}/(2\pi)$, where $B_{z_i}$ is the magnetic field projection along the i'th NV orientation, population is transferred to the $m_s=\pm1$ sublevel, resulting in greater population in the MS and lower probe transmission.

\begin{figure}
\centering
  \includegraphics[width=0.4\textwidth]{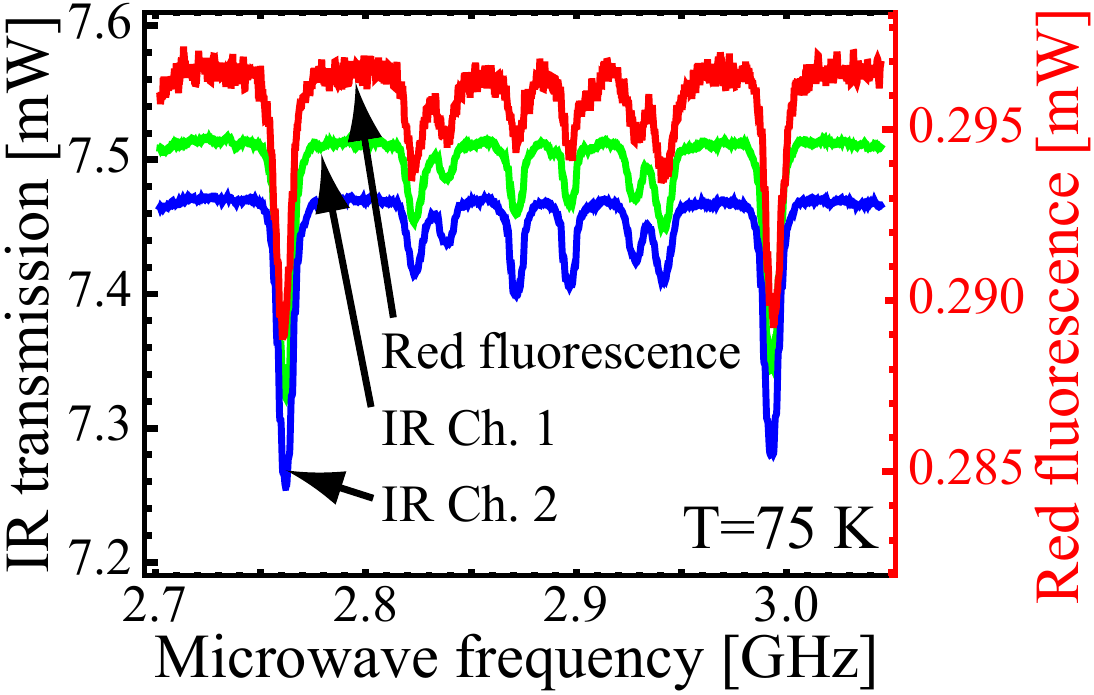}
\caption{\label{fig:magspec} \footnotesize  Optically-detected magnetic resonance at $75~$K using the fluorescence method and both halves of the transmitted IR probe. The pump power was $0.8~{\rm W}$ and the microwave Rabi frequency was $\sim1.5~{\rm MHz}$.}
\end{figure}

Magnetic-resonance spectra, detected by both fluorescence and IR-transmission, are shown in Fig. \ref{fig:magspec}. A bias magnetic field of $\sim4~{\rm mT}$, produced by a permanent magnet, was directed such that each of the four NV orientations had different $B_{z_i}$, resulting in eight resolved resonances. The contrast of both fluorescence and IR-transmission resonances depends on the change in the MS population, which saturates when the pump rate is $\Omega_{p}\gtrsim1/\tau_{MS}$, where $\tau_{MS}\approx0.3~{\rm \mu s}$ is the MS lifetime \cite{ACO2010ARXIV}. This condition was satisfied without significant power broadening for the pump power used here, $0.8~{\rm W}$.

For the IR-transmission resonances, the contrast also depends on the probe's optical depth. At room temperature, we find $R\sim0.003$, limited by the weak oscillator strength of the transition \cite{ROG2008,ACO2010ARXIV} and homogenous broadening of the IR absorption line \cite{ROG2008,ACO2010ARXIV}. The maximum contrast ($R\sim0.03$) occurs in the $45\mbox{-}75~{\rm K}$ temperature range, where the homogenous and inhomogenous contributions to the linewidth are approximately equal \cite{FU2009}. This contrast is, coincidentally, nearly the same as the maximum contrast obtained using fluorescence detection, but the signal is much larger due to the higher collected light intensity.

\begin{figure}
\centering
  \includegraphics[width=0.45\textwidth]{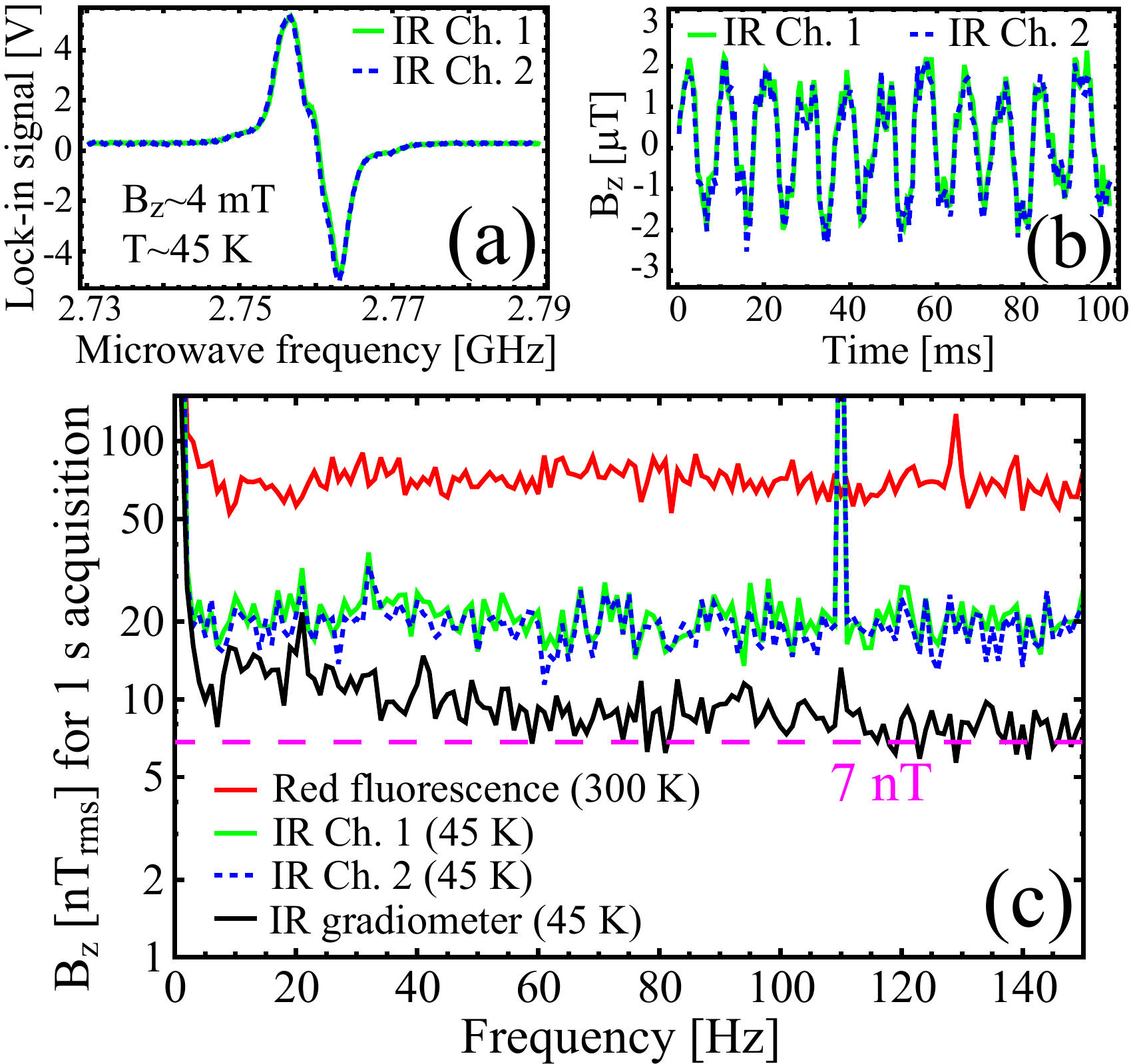}
\caption{\label{fig:opmag} \footnotesize  (a) Lock-in signal for both IR magnetometer channels. (b) Time series magnetometer signal, after subtraction of the static bias field, for a $1~{\rm \mu T_{rms}}$ applied field at 109 Hz. The microwave frequency was tuned to the center of the resonance (zero-crossing in (a)). (c) Frequency-domain response of the magnetometer output in (b) revealing an IR absorption gradiometer noise floor of $7~{\rm nT_{rms}}$ in 1 s of acquisition.}
\end{figure}

Operation of the device as a magnetometer was accomplished by phase-sensitive detection. An oscillating magnetic field (frequency 40 kHz, amplitude $\sim0.1~{\rm mT}$) was applied and the resulting photodiode signals were demodulated at the first harmonic using lock-in electronics. In order to maximize the contrast, we used the $\Delta m_s=-1$ resonance corresponding to NV orientation normal to the light polarization vectors \cite{ALE2007,ACO2010ARXIV}. Lock-in signals as a function of microwave frequency are shown in Fig. \ref{fig:opmag}(a) for both IR channels. The small difference in magnetic response of the two channels, due to inhomogeneity in the pump and microwave fields, was compensated for by adjusting the light levels hitting the photodiodes.

Tuning the microwaves to the center of the resonance, where the lock-in signal crosses zero, gave the highest magnetic response. Figure \ref{fig:opmag}(b) shows the time-series response of both magnetometer channels when an additional AC magnetic field (frequency 109 Hz, amplitude $1~{\rm \mu T_{rms}}$) was applied. The magnetometer was also operated at 26 and 426 Hz with a similar magnetic response. The measurement bandwidth in this work was limited by the lock-in time constant, but in principle this technique can be used to detect fields with angular frequency approaching the maximum spin polarization rate, $\sim1/\tau_{MS}$, without degradation in sensitivity.

The Fourier transform of the time-series response for the 109 Hz applied field, Fig. \ref{fig:opmag}(c), reveals a noise floor for each channel of $\sim15~{\rm nT_{rms}}$ in 1 s of acquisition for $\sim110~{\rm Hz}$ frequencies. For comparison, we also plot the noise spectrum for fluorescence-based magnetometer at room temperature, showing a noise floor of $\sim60~{\rm nT_{rms}}$ in 1 s of acquisition.

Taking the difference of the two IR channels' magnetometer signals gives a noise floor of $\sim7~{\rm nT_{rms}}$ in 1 s of acquisition for $\sim110~{\rm Hz}$ frequencies. Since these signals correspond to light that has interacted with spatially separate parts of the diamond, the difference-signal measures the magnetic field gradient across the beam (effective baseline $\sim25~{\rm \mu m}$). The benefit of this gradiometric approach is that technical noise common to both channels, such as laser intensity and ambient field fluctuations, is canceled. The cancelation here was imperfect, as the dominant contributions to the noise floor were uncompensated intensity fluctuations of both pump and probe lasers, as determined by blocking the pump laser and/or turning off the modulation field.

In the absence of technical noise, our technique is limited by photon shot-noise, given by:
\begin{equation}
\label{eq:photon}
\delta B_p \simeq \frac{1}{\gamma}\frac{\Gamma_{mr}}{R} \sqrt{\frac{E_p}{P t_m}},
\end{equation}
where $\Gamma_{mr}$ is the magnetic-resonance linewidth, $E_p$ is the photon energy, and $P$ is the detected optical power. Analyzing the highest-contrast resonances in Fig. \ref{fig:magspec}, we project a photon shot-noise limited sensitivity of $\sim40~{\rm pT}$ for fluorescence collection and $\sim5~{\rm pT}$ for the sum of both IR absorption channels at $t_m=1~{\rm s}$. The latter corresponds to approximately an order of magnitude better sensitivity-per-root-volume than the photon shot-noise limit of recent fluorescence-based demonstrations \cite{MAE2010,STE2010}.

Here we were limited by the available laser diode power, but if the 1042 nm transition is closed, we may be able to increase the probe power without adverse effects, making the photon shot-noise negligibly low, and leaving only the quantum-shot-noise-limited sensitivity. In this case, the number of NV centers that enters Eq. \eqref{eq:1} would be the change in the MS population on and off resonance.

The present technique can be extended to room-temperature operation by employing a cavity to increase the optical depth at 1042 nm. We estimate that a cavity with finesse of $\sim200$ will optimize the magnetometer response. Using two parallel micro-cavities should permit gradiometry with sensitivity approaching the quantum shot-noise limit of $\sim10~{\rm fT}$ in one second of acquisition. Such a device would be ideal for low-field NMR detection \cite{LED2009} in, for example, microfluidic devices \cite{LED2008PNAS}.

The authors thank B. Patton, E. Corsini, and the group of A. Pines for helpful comments. This work was supported by NSF.

\bibliographystyle{apsrev}

%\bibliography{diamond}

\begin{thebibliography}{22}
\expandafter\ifx\csname natexlab\endcsname\relax\def\natexlab#1{#1}\fi
\expandafter\ifx\csname bibnamefont\endcsname\relax
  \def\bibnamefont#1{#1}\fi
\expandafter\ifx\csname bibfnamefont\endcsname\relax
  \def\bibfnamefont#1{#1}\fi
\expandafter\ifx\csname citenamefont\endcsname\relax
  \def\citenamefont#1{#1}\fi
\expandafter\ifx\csname url\endcsname\relax
  \def\url#1{\texttt{#1}}\fi
\expandafter\ifx\csname urlprefix\endcsname\relax\def\urlprefix{URL }\fi
\providecommand{\bibinfo}[2]{#2}
\providecommand{\eprint}[2][]{\url{#2}}

\bibitem[{\citenamefont{Balasubramanian
  et~al.}(2008)\citenamefont{Balasubramanian, Chan, Kolesov, Al-Hmoud, Tisler,
  Shin, Kim, Wojcik, Hemmer, Krueger et~al.}}]{BAL2008}
\bibinfo{author}{\bibfnamefont{G.}~\bibnamefont{Balasubramanian}},
  \bibinfo{author}{\bibfnamefont{I.~Y.} \bibnamefont{Chan}},
  \bibinfo{author}{\bibfnamefont{R.}~\bibnamefont{Kolesov}},
  \bibinfo{author}{\bibfnamefont{M.}~\bibnamefont{Al-Hmoud}},
  \bibinfo{author}{\bibfnamefont{J.}~\bibnamefont{Tisler}},
  \bibinfo{author}{\bibfnamefont{C.}~\bibnamefont{Shin}},
  \bibinfo{author}{\bibfnamefont{C.}~\bibnamefont{Kim}},
  \bibinfo{author}{\bibfnamefont{A.}~\bibnamefont{Wojcik}},
  \bibinfo{author}{\bibfnamefont{P.~R.} \bibnamefont{Hemmer}},
  \bibinfo{author}{\bibfnamefont{A.}~\bibnamefont{Krueger}},
  \bibnamefont{et~al.}, \bibinfo{journal}{Nature}
  \textbf{\bibinfo{volume}{455}}, \bibinfo{pages}{648} (\bibinfo{year}{2008}).

\bibitem[{\citenamefont{Maze et~al.}(2008)\citenamefont{Maze, Stanwix, Hodges,
  Hong, Taylor, Cappellaro, Jiang, Dutt, Togan, Zibrov et~al.}}]{MAZ2008NATURE}
\bibinfo{author}{\bibfnamefont{J.~R.} \bibnamefont{Maze}},
  \bibinfo{author}{\bibfnamefont{P.~L.} \bibnamefont{Stanwix}},
  \bibinfo{author}{\bibfnamefont{J.~S.} \bibnamefont{Hodges}},
  \bibinfo{author}{\bibfnamefont{S.}~\bibnamefont{Hong}},
  \bibinfo{author}{\bibfnamefont{J.~M.} \bibnamefont{Taylor}},
  \bibinfo{author}{\bibfnamefont{P.}~\bibnamefont{Cappellaro}},
  \bibinfo{author}{\bibfnamefont{L.}~\bibnamefont{Jiang}},
  \bibinfo{author}{\bibfnamefont{M.~V.~G.} \bibnamefont{Dutt}},
  \bibinfo{author}{\bibfnamefont{E.}~\bibnamefont{Togan}},
  \bibinfo{author}{\bibfnamefont{A.~S.} \bibnamefont{Zibrov}},
  \bibnamefont{et~al.}, \bibinfo{journal}{Nature}
  \textbf{\bibinfo{volume}{455}}, \bibinfo{pages}{644} (\bibinfo{year}{2008}).

\bibitem[{\citenamefont{Taylor et~al.}(2008)\citenamefont{Taylor, Cappellaro,
  Childress, Jiang, Budker, Hemmer, Yacoby, Walsworth, and Lukin}}]{TAY2008}
\bibinfo{author}{\bibfnamefont{J.~M.} \bibnamefont{Taylor}},
  \bibinfo{author}{\bibfnamefont{P.}~\bibnamefont{Cappellaro}},
  \bibinfo{author}{\bibfnamefont{L.}~\bibnamefont{Childress}},
  \bibinfo{author}{\bibfnamefont{L.}~\bibnamefont{Jiang}},
  \bibinfo{author}{\bibfnamefont{D.}~\bibnamefont{Budker}},
  \bibinfo{author}{\bibfnamefont{P.~R.} \bibnamefont{Hemmer}},
  \bibinfo{author}{\bibfnamefont{A.}~\bibnamefont{Yacoby}},
  \bibinfo{author}{\bibfnamefont{R.}~\bibnamefont{Walsworth}},
  \bibnamefont{and} \bibinfo{author}{\bibfnamefont{M.~D.} \bibnamefont{Lukin}},
  \bibinfo{journal}{Nat Phys} \textbf{\bibinfo{volume}{4}},
  \bibinfo{pages}{810} (\bibinfo{year}{2008}).

\bibitem[{\citenamefont{Acosta et~al.}(2009)\citenamefont{Acosta, Bauch,
  Ledbetter, Santori, Fu, Barclay, Beausoleil, Linget, Roch, Treussart
  et~al.}}]{ACO2009}
\bibinfo{author}{\bibfnamefont{V.~M.} \bibnamefont{Acosta}},
  \bibinfo{author}{\bibfnamefont{E.}~\bibnamefont{Bauch}},
  \bibinfo{author}{\bibfnamefont{M.~P.} \bibnamefont{Ledbetter}},
  \bibinfo{author}{\bibfnamefont{C.}~\bibnamefont{Santori}},
  \bibinfo{author}{\bibfnamefont{K.~M.~C.} \bibnamefont{Fu}},
  \bibinfo{author}{\bibfnamefont{P.~E.} \bibnamefont{Barclay}},
  \bibinfo{author}{\bibfnamefont{R.~G.} \bibnamefont{Beausoleil}},
  \bibinfo{author}{\bibfnamefont{H.}~\bibnamefont{Linget}},
  \bibinfo{author}{\bibfnamefont{J.~F.} \bibnamefont{Roch}},
  \bibinfo{author}{\bibfnamefont{F.}~\bibnamefont{Treussart}},
  \bibnamefont{et~al.}, \bibinfo{journal}{Physical Review B}
  \textbf{\bibinfo{volume}{80}}, \bibinfo{pages}{115202}
  (\bibinfo{year}{2009}).

\bibitem[{\citenamefont{Bouchard et~al.}(2009)\citenamefont{Bouchard, Bauch,
  Acosta, and Budker}}]{BOU2010}
\bibinfo{author}{\bibfnamefont{L.~S.} \bibnamefont{Bouchard}},
  \bibinfo{author}{\bibfnamefont{E.}~\bibnamefont{Bauch}},
  \bibinfo{author}{\bibfnamefont{V.~M.} \bibnamefont{Acosta}},
  \bibnamefont{and} \bibinfo{author}{\bibfnamefont{D.}~\bibnamefont{Budker}},
  \emph{\bibinfo{title}{Detection of the meissner effect with a diamond
  magnetometer}} (\bibinfo{year}{2009}), \bibinfo{note}{arXiv:0911.2533v1
  [cond-mat.supr-con]}.

\bibitem[{\citenamefont{Steinert et~al.}(2010)\citenamefont{Steinert, Dolde,
  Neumann, Aird, Naydenov, Balasubramanian, Jelezko, and Wrachtrup}}]{STE2010}
\bibinfo{author}{\bibfnamefont{S.}~\bibnamefont{Steinert}},
  \bibinfo{author}{\bibfnamefont{F.}~\bibnamefont{Dolde}},
  \bibinfo{author}{\bibfnamefont{P.}~\bibnamefont{Neumann}},
  \bibinfo{author}{\bibfnamefont{A.}~\bibnamefont{Aird}},
  \bibinfo{author}{\bibfnamefont{B.}~\bibnamefont{Naydenov}},
  \bibinfo{author}{\bibfnamefont{G.}~\bibnamefont{Balasubramanian}},
  \bibinfo{author}{\bibfnamefont{F.}~\bibnamefont{Jelezko}}, \bibnamefont{and}
  \bibinfo{author}{\bibfnamefont{J.}~\bibnamefont{Wrachtrup}},
  \bibinfo{journal}{Review of Scientific Instruments}
  \textbf{\bibinfo{volume}{81}}, \bibinfo{pages}{043705}
  (\bibinfo{year}{2010}).

\bibitem[{\citenamefont{Maertz et~al.}(2010)\citenamefont{Maertz, Wijnheijmer,
  Fuchs, Nowakowski, and Awschalom}}]{MAE2010}
\bibinfo{author}{\bibfnamefont{B.~J.} \bibnamefont{Maertz}},
  \bibinfo{author}{\bibfnamefont{A.~P.} \bibnamefont{Wijnheijmer}},
  \bibinfo{author}{\bibfnamefont{G.~D.} \bibnamefont{Fuchs}},
  \bibinfo{author}{\bibfnamefont{M.~E.} \bibnamefont{Nowakowski}},
  \bibnamefont{and} \bibinfo{author}{\bibfnamefont{D.~D.}
  \bibnamefont{Awschalom}}, \bibinfo{journal}{Applied Physics Letters}
  \textbf{\bibinfo{volume}{96}}, \bibinfo{pages}{092504}
  (\bibinfo{year}{2010}).

\bibitem[{\citenamefont{Rogers et~al.}(2008)\citenamefont{Rogers, Armstrong,
  Sellars, and Manson}}]{ROG2008}
\bibinfo{author}{\bibfnamefont{L.~G.} \bibnamefont{Rogers}},
  \bibinfo{author}{\bibfnamefont{S.}~\bibnamefont{Armstrong}},
  \bibinfo{author}{\bibfnamefont{M.~J.} \bibnamefont{Sellars}},
  \bibnamefont{and} \bibinfo{author}{\bibfnamefont{N.~B.}
  \bibnamefont{Manson}}, \bibinfo{journal}{New Journal of Physics}
  \textbf{\bibinfo{volume}{10}}, \bibinfo{pages}{103024}
  (\bibinfo{year}{2008}).

\bibitem[{\citenamefont{Acosta et~al.}(2010{\natexlab{a}})\citenamefont{Acosta,
  Jarmola, Bauch, and Budker}}]{ACO2010ARXIV}
\bibinfo{author}{\bibfnamefont{V.~M.} \bibnamefont{Acosta}},
  \bibinfo{author}{\bibfnamefont{A.}~\bibnamefont{Jarmola}},
  \bibinfo{author}{\bibfnamefont{E.}~\bibnamefont{Bauch}}, \bibnamefont{and}
  \bibinfo{author}{\bibfnamefont{D.}~\bibnamefont{Budker}},
  \emph{\bibinfo{title}{Optical properties of the nitrogen-vacancy singlet
  levels in diamond}} (\bibinfo{year}{2010}{\natexlab{a}}),
  \bibinfo{note}{arXiv:1009.0032v1 [quant-ph]}.

\bibitem[{\citenamefont{Budker and Romalis}(2007)}]{BUD2007}
\bibinfo{author}{\bibfnamefont{D.}~\bibnamefont{Budker}} \bibnamefont{and}
  \bibinfo{author}{\bibfnamefont{M.}~\bibnamefont{Romalis}},
  \bibinfo{journal}{Nature Physics} \textbf{\bibinfo{volume}{3}},
  \bibinfo{pages}{227} (\bibinfo{year}{2007}).

\bibitem[{\citenamefont{Loubser and van Wyk}(1978)}]{LOU1978}
\bibinfo{author}{\bibfnamefont{J.}~\bibnamefont{Loubser}} \bibnamefont{and}
  \bibinfo{author}{\bibfnamefont{J.~A.} \bibnamefont{van Wyk}},
  \bibinfo{journal}{Reports on Progress in Physics}
  \textbf{\bibinfo{volume}{41}}, \bibinfo{pages}{1201} (\bibinfo{year}{1978}).

\bibitem[{\citenamefont{van Oort et~al.}(1988)\citenamefont{van Oort, Manson,
  and Glasbeek}}]{VAN1988}
\bibinfo{author}{\bibfnamefont{E.}~\bibnamefont{van Oort}},
  \bibinfo{author}{\bibfnamefont{N.~B.} \bibnamefont{Manson}},
  \bibnamefont{and} \bibinfo{author}{\bibfnamefont{M.}~\bibnamefont{Glasbeek}},
  \bibinfo{journal}{Journal of Physics C-Solid State Physics}
  \textbf{\bibinfo{volume}{21}}, \bibinfo{pages}{4385} (\bibinfo{year}{1988}).

\bibitem[{\citenamefont{Jelezko and Wrachtrup}(2006)}]{JEL2006}
\bibinfo{author}{\bibfnamefont{F.}~\bibnamefont{Jelezko}} \bibnamefont{and}
  \bibinfo{author}{\bibfnamefont{J.}~\bibnamefont{Wrachtrup}},
  \bibinfo{journal}{Physica Status Solidi a-Applications and Materials Science}
  \textbf{\bibinfo{volume}{203}}, \bibinfo{pages}{3207} (\bibinfo{year}{2006}).

\bibitem[{\citenamefont{Manson et~al.}(2006)\citenamefont{Manson, Harrison, and
  Sellars}}]{MAN2006}
\bibinfo{author}{\bibfnamefont{N.~B.} \bibnamefont{Manson}},
  \bibinfo{author}{\bibfnamefont{J.~P.} \bibnamefont{Harrison}},
  \bibnamefont{and} \bibinfo{author}{\bibfnamefont{M.~J.}
  \bibnamefont{Sellars}}, \bibinfo{journal}{Physical Review B}
  \textbf{\bibinfo{volume}{74}}, \bibinfo{pages}{104303}
  (\bibinfo{year}{2006}).

\bibitem[{\citenamefont{Alegre et~al.}(2007)\citenamefont{Alegre, Santori,
  Medeiros-Ribeiro, and Beausoleil}}]{ALE2007}
\bibinfo{author}{\bibfnamefont{T.~P.~M.} \bibnamefont{Alegre}},
  \bibinfo{author}{\bibfnamefont{C.}~\bibnamefont{Santori}},
  \bibinfo{author}{\bibfnamefont{G.}~\bibnamefont{Medeiros-Ribeiro}},
  \bibnamefont{and} \bibinfo{author}{\bibfnamefont{R.~G.}
  \bibnamefont{Beausoleil}}, \bibinfo{journal}{Physical Review B}
  \textbf{\bibinfo{volume}{76}}, \bibinfo{pages}{165205}
  (\bibinfo{year}{2007}).

\bibitem[{\citenamefont{Han et~al.}(2010)\citenamefont{Han, Kim, Eggeling, and
  Hell}}]{HAN2010}
\bibinfo{author}{\bibfnamefont{K.~Y.} \bibnamefont{Han}},
  \bibinfo{author}{\bibfnamefont{S.~K.} \bibnamefont{Kim}},
  \bibinfo{author}{\bibfnamefont{C.}~\bibnamefont{Eggeling}}, \bibnamefont{and}
  \bibinfo{author}{\bibfnamefont{S.~W.} \bibnamefont{Hell}},
  \bibinfo{journal}{Nano Letters} \textbf{\bibinfo{volume}{10}},
  \bibinfo{pages}{3199} (\bibinfo{year}{2010}).

\bibitem[{\citenamefont{Babinec et~al.}(2010)\citenamefont{Babinec, Hausmann,
  Khan, Zhang, Maze, Hemmer, and Loncar}}]{BAB2010}
\bibinfo{author}{\bibfnamefont{T.~M.} \bibnamefont{Babinec}},
  \bibinfo{author}{\bibfnamefont{B.~J.~M.} \bibnamefont{Hausmann}},
  \bibinfo{author}{\bibfnamefont{M.}~\bibnamefont{Khan}},
  \bibinfo{author}{\bibfnamefont{Y.}~\bibnamefont{Zhang}},
  \bibinfo{author}{\bibfnamefont{J.~R.} \bibnamefont{Maze}},
  \bibinfo{author}{\bibfnamefont{P.~R.} \bibnamefont{Hemmer}},
  \bibnamefont{and} \bibinfo{author}{\bibfnamefont{M.}~\bibnamefont{Loncar}},
  \bibinfo{journal}{Nat Nano} \textbf{\bibinfo{volume}{5}},
  \bibinfo{pages}{195} (\bibinfo{year}{2010}).

\bibitem[{\citenamefont{Acosta et~al.}(2010{\natexlab{b}})\citenamefont{Acosta,
  Bauch, Ledbetter, Waxman, Bouchard, and Budker}}]{ACO2010}
\bibinfo{author}{\bibfnamefont{V.~M.} \bibnamefont{Acosta}},
  \bibinfo{author}{\bibfnamefont{E.}~\bibnamefont{Bauch}},
  \bibinfo{author}{\bibfnamefont{M.~P.} \bibnamefont{Ledbetter}},
  \bibinfo{author}{\bibfnamefont{A.}~\bibnamefont{Waxman}},
  \bibinfo{author}{\bibfnamefont{L.~S.} \bibnamefont{Bouchard}},
  \bibnamefont{and} \bibinfo{author}{\bibfnamefont{D.}~\bibnamefont{Budker}},
  \bibinfo{journal}{Physical Review Letters} \textbf{\bibinfo{volume}{104}},
  \bibinfo{pages}{070801} (\bibinfo{year}{2010}{\natexlab{b}}).

\bibitem[{\citenamefont{Balasubramanian
  et~al.}(2009)\citenamefont{Balasubramanian, Neumann, Twitchen, Markham,
  Kolesov, Mizuochi, Isoya, Achard, Beck, Tissler et~al.}}]{BAL2009}
\bibinfo{author}{\bibfnamefont{G.}~\bibnamefont{Balasubramanian}},
  \bibinfo{author}{\bibfnamefont{P.}~\bibnamefont{Neumann}},
  \bibinfo{author}{\bibfnamefont{D.}~\bibnamefont{Twitchen}},
  \bibinfo{author}{\bibfnamefont{M.}~\bibnamefont{Markham}},
  \bibinfo{author}{\bibfnamefont{R.}~\bibnamefont{Kolesov}},
  \bibinfo{author}{\bibfnamefont{N.}~\bibnamefont{Mizuochi}},
  \bibinfo{author}{\bibfnamefont{J.}~\bibnamefont{Isoya}},
  \bibinfo{author}{\bibfnamefont{J.}~\bibnamefont{Achard}},
  \bibinfo{author}{\bibfnamefont{J.}~\bibnamefont{Beck}},
  \bibinfo{author}{\bibfnamefont{J.}~\bibnamefont{Tissler}},
  \bibnamefont{et~al.}, \bibinfo{journal}{Nat Mater}
  \textbf{\bibinfo{volume}{8}}, \bibinfo{pages}{383} (\bibinfo{year}{2009}).

\bibitem[{\citenamefont{Fu et~al.}(2009)\citenamefont{Fu, Santori, Barclay,
  Rogers, Manson, and Beausoleil}}]{FU2009}
\bibinfo{author}{\bibfnamefont{K.-M.~C.} \bibnamefont{Fu}},
  \bibinfo{author}{\bibfnamefont{C.}~\bibnamefont{Santori}},
  \bibinfo{author}{\bibfnamefont{P.~E.} \bibnamefont{Barclay}},
  \bibinfo{author}{\bibfnamefont{L.~J.} \bibnamefont{Rogers}},
  \bibinfo{author}{\bibfnamefont{N.~B.} \bibnamefont{Manson}},
  \bibnamefont{and} \bibinfo{author}{\bibfnamefont{R.~G.}
  \bibnamefont{Beausoleil}}, \bibinfo{journal}{Physical Review Letters}
  \textbf{\bibinfo{volume}{103}}, \bibinfo{pages}{256404}
  (\bibinfo{year}{2009}).

\bibitem[{\citenamefont{Ledbetter et~al.}(2009)\citenamefont{Ledbetter,
  Crawford, Pines, Wemmer, Knappe, Kitching, and Budker}}]{LED2009}
\bibinfo{author}{\bibfnamefont{M.~P.} \bibnamefont{Ledbetter}},
  \bibinfo{author}{\bibfnamefont{C.~W.} \bibnamefont{Crawford}},
  \bibinfo{author}{\bibfnamefont{A.}~\bibnamefont{Pines}},
  \bibinfo{author}{\bibfnamefont{D.~E.} \bibnamefont{Wemmer}},
  \bibinfo{author}{\bibfnamefont{S.}~\bibnamefont{Knappe}},
  \bibinfo{author}{\bibfnamefont{J.}~\bibnamefont{Kitching}}, \bibnamefont{and}
  \bibinfo{author}{\bibfnamefont{D.}~\bibnamefont{Budker}},
  \bibinfo{journal}{Journal of Magnetic Resonance}
  \textbf{\bibinfo{volume}{199}}, \bibinfo{pages}{25} (\bibinfo{year}{2009}).

\bibitem[{\citenamefont{Ledbetter et~al.}(2008)\citenamefont{Ledbetter,
  Savukov, Budker, Shah, Knappe, Kitching, Michalak, Xu, and
  Pines}}]{LED2008PNAS}
\bibinfo{author}{\bibfnamefont{M.~P.} \bibnamefont{Ledbetter}},
  \bibinfo{author}{\bibfnamefont{I.~M.} \bibnamefont{Savukov}},
  \bibinfo{author}{\bibfnamefont{D.}~\bibnamefont{Budker}},
  \bibinfo{author}{\bibfnamefont{V.}~\bibnamefont{Shah}},
  \bibinfo{author}{\bibfnamefont{S.}~\bibnamefont{Knappe}},
  \bibinfo{author}{\bibfnamefont{J.}~\bibnamefont{Kitching}},
  \bibinfo{author}{\bibfnamefont{D.~J.} \bibnamefont{Michalak}},
  \bibinfo{author}{\bibfnamefont{S.}~\bibnamefont{Xu}}, \bibnamefont{and}
  \bibinfo{author}{\bibfnamefont{A.}~\bibnamefont{Pines}},
  \bibinfo{journal}{Proceedings of the National Academy of Sciences of the
  United States of America} \textbf{\bibinfo{volume}{105}},
  \bibinfo{pages}{2286} (\bibinfo{year}{2008}).

\end{thebibliography}
\end{document}